\documentclass[10pt]{article}
\usepackage{graphics,epsfig}

\begin{document}
\title{Why the Disjunction in Quantum Logic is Not
Classical\footnote{To appear in Foundations of Physics, volume {\bf 30},
issue 10, 2000.}}
\author{Diederik Aerts, Ellie D'Hondt, and Liane Gabora}
\date{}
\maketitle
\centerline {Center Leo Apostel, Brussels Free University}
\centerline {Krijgskundestraat 33, 1160 Brussels, Belgium.}
\centerline {diraerts@vub.ac.be, eldhondt@vub.ac.be, lgabora@vub.ac.be}
\begin{abstract}
\noindent
In this paper, the quantum logical `or' is analyzed from a physical
perspective. We show that it is the existence of EPR-like correlation states
for the quantum mechanical entity under consideration that make it
nonequivalent to the classical situation. Specifically, the presence of
potentiality in these correlation states gives rise to the quantum
deviation from the classical logical
`or'. We show how this arises not only in the microworld, but also in
macroscopic situations where EPR-like correlation states are
present. We investigate how application of this analysis to concepts could
alleviate some well known
problems in cognitive science.
\end{abstract}
\bigskip
\noindent
{\it Dedication} We dedicate this paper to Marisa Dalla Chiara, one of the
founding figures of quantum logic. The ideas expressed
in this article have been influenced by her ground-breeaking work in this
field.
\section{Introduction}
We put forward a physical explanation of why, in quantum logic, the logical
disjunction does not
behave classically, even for compatible propositions. Most studies of
quantum logic have
concentrated on the algebraic structure of the set of propositions, trying
to identify the structural differences
between quantum logic and classical logic (von Neumann 1932; Birkhoff and
von Neumann 1936; Beltrametti and Cassinelli
1981). These mathematical studies, have shown that the quantum logical
implication and conjunction can be
interpreted as their classical equivalents, while this is not the case for
the quantum logical disjunction and negation. We
will show that it is the presence of EPR-type quantum mechanical
correlations that is at the origin of the
nonclassical behavior of the logical disjunction.

In quantum logic, a proposition $a$ is represented by means
of the closed subspace $M_a$ of the Hilbert space ${\cal H}$ used to
describe the quantum entity under consideration,
or by means of the orthogonal projection operator
$P_a$ on this closed subspace. We will use both representations, since some
of the logical relations in
quantum logic can be more easily expressed using the `closed subspace'
representation of a proposition, while others
are more easily expressed using the `orthogonal projection' representation.
Let us denote the set of propositions of the
quantum entity under consideration by means of ${\cal P}$, the set of
closed subspaces of the Hilbert space ${\cal H}$,
describing the quantum entity by means of
${\cal L}({\cal H})$ and the set of orthogonal projection operators by means of
${\cal P}({\cal H})$. A state $p$ of the quantum mechanical entity under
consideration is represented by means of the
unit vector $v_p$ of the Hilbert space ${\cal H}$.

\section{The Quantum Logical Operations}

For two propositions $a, b \in {\cal P}$ the quantum logical operations are
introduced by the following expressions:
\begin{eqnarray}
a \mapsto b &\Leftrightarrow& M_a \subset M_b  \label{implication} \\
M_{a \wedge b} &=& M_a \cap M_b \label{conjunction} \\
M_{a \vee b} &=& cl(M_a \cup M_b) \\
M_{\neg a} &=& M_a^\perp
\end{eqnarray}
We remark that $cl(M_a \cup M_b)$ is the topological closure of the linear
space generated by $M_a \cup M_b$. This
means that it is the smallest closed subspace of ${\cal H}$ that contains
$M_a$ and $M_b$.

Using these standard
definitions of the quantum logical operations, we can retrieve the `truth' and
`falseness' of the various
possibilities. Suppose that proposition $a \in {\cal P}$ is true. This
means that the state $p$ of the quantum
entity is such that whenever $a$ undergoes a `yes-no' test
$\alpha$, the outcome `yes' can be predicted with certainty (probability
equal to 1). As we know, such a `yes-no' test in quantum
mechanics is represented by the self adjoint operator, the spectral
decomposition of which is given by the orthogonal
projections $P_a$ and $I - P_a$, where $I$ is the unit operator of the
Hilbert space ${\cal H}$. From the formalism
of quantum mechanics, it follows that proposition $a \in {\cal P}$ is true
iff the state $p$ of the quantum entity
is such that $P_av_p = v_p$, which is equivalent to $v_p \in M_a$.

\subsection{The Implication}

Let us now consider the quantum logical implication. Suppose we have two
propositions $a, b \in {\cal P}$ such that $a
\mapsto b$, and suppose that $a$ is true. This means that the quantum
mechanical entity under consideration is such
that for its state $p$ we have $v_p \in M_a$. Since from equation
\ref{implication} it follows that $M_a \subset
M_b$, we have $v_p \in M_b$. This shows that also $b$ is true. This in turn
shows that the meaning of $a \mapsto b$
is the following: `if $a$ is true, then it follows that $b$ is true'. As a
consequence, the quantum logical implication
behaves in the same way as the classical logical implication.

\subsection{The Conjunction}
Let us consider the quantum logical conjunction. For two propositions $a, b
\in {\cal P}$ we consider $a \wedge b$ to be
true. This means that the state $p$ of the quantum entity under
consideration is such that $v_p \in M(a \wedge b)$.
>From equation \ref{conjunction}, it follows that this is equivalent to $v_p
\in M_a \cap M_b$, which is equivalent to
$v_p \in M_a$ `and' $v_p \in M_a$. This again is equivalent to $a$ is true
`and' $b$ is true. Thus we have shown
that $a \wedge b$ is true $\Leftrightarrow$ $a$ true `and' $b$ true. As a
consequence, the quantum logical conjunction
behaves in the same way as the classical logical conjunction.

\subsection{The Disjunction}
Now we will consider the quantum logical disjunction. For two propositions
$a, b \in {\cal P}$, we consider $a \vee b$.
Let us find out when $a \vee b$ is true. We remark that $M_a \subset cl(M_a
\cup M_b)$ and $M_b \subset cl(M_a \cup
M_b)$, which shows that $a \mapsto a \vee b$ and $b \mapsto a \vee b$. This
means that if $a$ `or' $b$ is true it
follows that $a \vee b$ is true. The inverse implication, however, does not
hold. Indeed, $a \vee b$ can be true
without $a$ `or' $b$ being true. The reason is that $cl(M_a \cup M_b)$
contains, in general, vectors that are not
contained in $M_a$ or $M_b$. If the quantum entity is in a state $p$ where
$v_p$ is such a vector, then $a \vee
b$ is true without $a$ or $b$ being true. This shows that the disjunction
in quantum logic cannot be interpreted as the
disjunction of classical logic.

\subsection{The Negation}
Although it is not the subject of this paper, we can easily see that the
quantum logic negation is also not the
same as the classical logic negation. Indeed, consider a proposition $a \in
{\cal P}$ and suppose that $\neg a$ is true.
This means that the state $p$ of the considered quantum entity is such that
$v_p \in M_a^\perp$. Since $M_a^\perp
\cap M_a = \emptyset$ we have that $v_p \notin M_a$, and hence $a$ is not
true. This means that if the quantum
negation of a proposition is true, then the classical negation of this
proposition is true. However the inverse does not
hold. In other words, it is possible that the quantum entity is in a state
such that $a$ is not true, without $\neg a$
being true.

\section{EPR-like Correlations and the Nonclassical Nature of Disjunction}

As we have shown in the foregoing section, the reason the quantum logical
disjunction does not behave classically
is that, for two propositions $a, b \in {\cal P}$, the quantum entity can
be in a state $p$, such that $a \vee b$ is
true without $a$ being true or $b$ being true. For such a state $p$ we have that
$v_p \in cl(M_a \cup M_b)$, but
$v_p \notin M_a$ and $v_p \notin M_b$. We now put forward the main result
of this paper: the
presence of EPR-like correlations is the origin of the nonclassical nature
of the quantum disjunction for the case of compatible
propositions.

\subsection{Compatible Propositions and Truth Tables}

Let us now consider two propositions $a, b \in {\cal P}$ that are
compatible, which means that $P_aP_b = P_bP_a$. In this case,
the two `yes-no' experiments $\alpha$ and $\beta$ that test $a$ and $b$ can
be performed together. The experiment
that consists of testing the two propositions together, which we denote
$\alpha \wedge \beta$, has four possible
outcomes (yes, yes), (yes, no), (no, yes) and (no, no). In classical logic,
the logical operations can be defined by means of
truth tables, and for compatible quantum propositions we can also introduce
truth tables. Considering the experiment $\alpha
\wedge \beta$ we say that the conjunction $a \wedge b$ is true iff the
state of the quantum entity is such that for the
experiment $\alpha \wedge
\beta$ we obtain with certainty the outcome (yes, yes). Similarly, we say
that the disjunction $a \vee b$ is true iff the state
of the quantum entity is such that for the experiment $\alpha \wedge \beta$
we obtain with certainty one of the outcomes (yes, yes),
(yes, no) or (no, yes).

\subsection{Compatible Propositions and EPR-like Correlations}
Suppose now that we are in a situation where EPR-type correlations exist in
relation to the two propositions $a$ and $b$.
This means that the state of the quantum entity is such that the
measurement $\alpha \wedge \beta$ always leads to the
outcome (yes, no) or (no, yes). As a consequence, $a \vee b$ is true. But
is is clear that neither $a$ nor $b$ are true in
general, which shows that $a$ `or' $b$ is not true. It is the possibility
of such a correlated EPR state that
makes the quantum logical disjunction differ from the classical logical
disjunction.

\section{Construction of an EPR-like State for a Quantum Entity}
In this section we show that the EPR-like state can be constructed by means
of the superposition principle for any two compatible
propositions.

Consider $a, b \in {\cal P}$ compatible propositions of a quantum entity
described in a Hilbert space ${\cal H}$, such that
$P_a(1-P_b)({\cal H}) \not= \emptyset$ and $(1 - P_a)P_b({\cal H}) \not=
\emptyset$. A self adjoint operator that corresponds to
the measurement of the experiment
$\alpha
\wedge
\beta$ is given by:

\begin{equation}
H = \lambda_1 P_a P_b + \lambda_2 P_a (1 - P_b) + \lambda_3 (1 - P_a) P_b +
\lambda_4 (1 - P_a)(1 - P_b)
\end{equation}
where $\lambda_1, \lambda_2, \lambda_3$ and $\lambda_4$ are real numbers.
The values $\lambda_1, \lambda_2, \lambda_3$ and
$\lambda_4$ correspond respectively to the outcomes (yes, yes), (yes, no),
(no, yes) and (no, no) of the experiment $\alpha
\wedge \beta$. Consider unit vectors
$x, y
\in {\cal H}$ such that
$P_a(1-P_b) x = x$ and $(1 - P_a)P_b y = y$.

We have
\begin{equation}
P_aP_b x = (1-P_a)P_b x = (1-P_a)(1-P_b)x = 0
\end{equation}
\begin{equation}
P_aP_b y = P_a(1-P_b) y = (1-P_a)(1-P_b) y = 0
\end{equation}
Let us indicate how these equalities are derived. For example $P_aP_b x =
P_aP_bP_a(1-P_b)x = P_aP_b(1-P_b)x = 0$. The other
equalities are derived in an analogous way.
Let us consider a state $p$ of the quantum entity such that
\begin{equation}
v_p = {1 \over \sqrt2}(x + y)
\end{equation}
We have
\begin{equation}
P_aP_bv_p = (1-P_a)(1-P_b)v_p = 0
\end{equation}
\begin{equation}
P_a(1-P_b)v_p = {1 \over \sqrt2}x
\end{equation}
\begin{equation}
(1-P_a)P_bv_p = {1 \over \sqrt2}y
\end{equation}
Using the quantum formalism and the just derived formulas we can calculate
the probabilities, if the quantum
entity is in state $v_p$, for a measurement $\alpha
\wedge \beta$ to lead to the different outcomes (yes, yes), (yes, no), (no,
yes) and (no, no), let us denote them respectively
$\mu(\alpha \wedge \beta, yes, yes), \mu(\alpha \wedge \beta, yes, no), \mu(\alpha \wedge \beta, no, yes)$ and $\mu(\alpha
\wedge \beta, no, no)$. We have
\begin{equation}
\mu(\alpha \wedge \beta, yes, yes) = <P_aP_bv_p, P_aP_bv_p> = 0
\end{equation}
\begin{equation}
\mu(\alpha \wedge \beta, yes, no) = <P_a(1-P_b)v_p, P_a(1-P_b)v_p> = {1
\over 2}
\end{equation}
\begin{equation}
\mu(\alpha \wedge \beta, no, yes) = <(1-P_a)P_bv_p, (1-P_a)P_bv_p> = {1
\over 2}
\end{equation}
\begin{equation}
\mu(\alpha \wedge \beta, no, no) = <(1-P_a)(1-P_b)v_p, (1-P_a)(1-P_b)v_p> = 0
\end{equation}
This proves that for the quantum entity being in state $p$ the experiment
$\alpha \wedge \beta$ gives rise to EPR-like
correlations for the propositions $a$ and $b$. The possible outcomes are
(yes, no) or (no, yes).

\section{The Quantum `Or' in the Macroscopic World}
Elsewhere in this volume, is a paper that shows that EPR-like correlations
also exist for macroscopic entities,
depending on the state and propositions that are considered (Aerts et al.,
2000). The examples in that paper also shed light on the
current subject, so we touch on them again here briefly. For the `connected
vessels of water' example we consider two propositions
$a$ and
$b$. Proposition $a$ is defined by the sentence: `there is more than 10
liters of water at the left', and
proposition $b$ by the sentence: `there is more than 10 liters of water at
the right'.

The measurement $\alpha$ tests
proposition $a$ by pouring out the water at the left with a siphon, and
collecting it in a reference vessel, and the measurement
$\beta$ does the same at the right. If we test proposition $a$ for the
state of the connected vessels containing 20
liters of water, we find that $\alpha$ gives the outcome `yes' with
certainty, and also $\beta$ gives the outcome `yes' with
certainty. If we test the propositions separately, we pour the whole 20
liters out at the left side as well as the right. At first
sight this seems to suggest that both propositions are true at once and
hence that $a \wedge b$ is true. But after getting a better
look we see that this is an error. Indeed, obviously when
we pour out the water at the left it influences what happens to the water
at the right. More concretely the water at right is also
poured out, and hence helps to result in there being more than 10 liters at
the left. Indeed, we also know that there cannot be more
than 10 liters of water to left and more than 10 liters of water to the
right, because the total must equal 20. Our error was to
believe that we can test propositions separately in this situation. So let
us correct this error by introducing the measurement
$\alpha \wedge \beta$ that tests the two propositions together, by pouring
out the water at both sides at once. The result is
then that if we have more than 10 liters at the left, we have less than 10
liters at the right, and if we have more than 10
liters at the right, we have less than 10 liters at the left. This means
that $a \wedge b$ is certainly not true. On the contrary,
each time we find $a$ to be true, $b$ turns out not to be true, and vice
versa. However, $a \vee b$ is
still true, since for $\alpha \wedge \beta$ we always have one of the
outcomes (yes, no) `or' (no, yes). Would this then mean that
$a$ `or' $b$ is true, or equivalently $a$ is true `or' $b$ is true?
Definitely not.

Indeed if $a$ is true `or' $b$ is true, the measurement $\alpha \wedge
\beta$ should give with certainty (yes, yes) or (yes,
no), in which case
$a$ is true, `or' it should give with certainty (yes, yes) or (no, yes), in
which case $b$ is true. The real
situation is more subtle. The connected vessels of water potentially
contain `more than
10 liters of water to the left' `or' `more than 10 liters of water to the
right', but this potentiality is not made
actual before the measurement $\alpha \wedge \beta$ is finished. This is
expressed by stating that
proposition $a \vee b$ is true. It also shows that $a \vee b$ is not
equivalent to $a$ `or' $b$ as a proposition. It is
the possibility of the potential state of the connected vessels of water
that makes the `or'proposition nonclassical.

Let us now turn to the example from cognitive science treated in Aerts et
al. 2000. We could restate the insight of the
foregoing paragraph for the case of concepts in the mind as follows. We
introduce the set of propositions $\{a_n\}$,
where $a_n$ is the proposition `the mind thinks of instance $n$', where
each $n$ is an instance of the concept
`cat'. Suppose that the state of the mind is such that it thinks of the
concept `cat'. Just as with
the vessels of water, we can say that one of the propositions $a_n$ is
true, but only potentially. This is again
expressed by the proposition
$a_1 \vee a_2 \vee ... \vee a_i \vee ... \vee a_n$ not being equivalent to
the proposition $a_1$ `or' $a_2$ `or' ... `or'
$a_i$ `or' ... `or' $a_n$. Thus we cannot describe a concept as simply a
set of instances.
It differs from the instances in the same way the connected vessels
containing 20 liters of water is
different from the set of all separated vessels with water summming to 20
liters. This difference is identical to
the well known difference between the electron as described by modern
quantum mechanics, and the model that
was proposed for the electron in the old quantum theory (before 1926) of a
`cloud' of charged particles inside the atom.

We are currently analyzing how this approach to concepts can shed light on
well known
problems in cognitive science such as the `pet fish
problem'. Experimental research shows that `guppy' is not a good example of
the concept
`pet', nor is it a good example of the concept `fish', but it is indeed a
good example of the concept `pet fish' (Osherson and
Smith, 1981). Prelimenary investigation indicates that many of the problems
that arise with other formal approaches
to conceptual dynamics (see Rosch 2000 for a summary) can be resolved using
a quantum mechanical approach.

\section{Conclusion}
We have shown how the quantum logical `or' and its nonequivalence with the
classical logical
`or' can be understood from a physical perspective. The origin of the
quantum logical `or'  and its
difference with the classical logical `or', is the presence
of `potential correlations' of the EPR-type.

This type of potentiality does not only appear in the
microworld, where it is abundant, but also in the macroworld. Special
attention has been given to application of this insight to concepts in the
mind.

\section{References}

\medskip
\noindent
Aerts, D., Aerts, S., Broekaert, J. and Gabora, L., 2000, ``The violation
of Bell inequalities in the macroworld", Foundations
of Physics, this issue.

\medskip
\noindent
Beltrametti, E. G. and Cassinelli, J., 1981, {\it The Logic of Quantum
Mechanics}, Addison Wesley, Reading.

\medskip
\noindent
Birkhoff, G. and von Neumann, J., 1936, ``The logic of quantum mechanics",
{\it Annals of Mathematics}, {\bf 43}, 298 -
331.

\medskip
\noindent
Osherson, D. N. and Smith, E. E., 1981, ``On the adequacy of prototype
theory as a theory of concepts", Cognition, {\bf
9}, 35 -58.

\medskip
\noindent
Rosch, E., 2000, ``Reclaiming concepts", Journal of Consciousness Studies,
{\bf 6}, 61 -77.

\medskip
\noindent
von Neumann, J., 1932, {\it Mathematische Grundlagen der Quantenmechanik},
Springer-Verlag, Berlin.

\end{document}